\documentclass[aps,prb,twocolumn,showpacs,amsmath,amssymb,superscriptaddress,floatfix]{revtex4}
\usepackage{graphicx}
\usepackage{Jonasmacros}
\usepackage{pstricks}

\begin{document}
\preprint{\today}
\title{Pauli spin blockade in weakly coupled double quantum dots}
\author{J. Fransson}
\email{Jonas.Fransson@fysik.uu.se}
\affiliation{Department of Materials Science and Engineering, Royal Institute of Technology (KTH), SE-100 44\ \ Stockholm, Sweden}
\affiliation{Physics Department, Uppsala University, Box 530, SE-751 21\ \ Uppsala, Sweden}
\author{M. R\aa sander}
\affiliation{Physics Department, Uppsala University, Box 530, SE-751 21\ \ Uppsala, Sweden}

\begin{abstract}
In a two-level system, constituted by two serially coupled single level quantum dots, coupled to external leads we find that the current is suppressed in one direction of biasing caused by a fully occupied two-electron triplet state in the interacting region. The efficiency of the current suppression is governed by the ratio between the interdot tunnelling rate and the level off-set. In the opposite bias direction, the occupation of the two-electron triplet is lifted which allows a larger current to flow through the system, where the conductance is provided by transitions between one-electron states and two-electron singlet states. Is is also shown that a finite ferromagnetic interdot exchange interaction provides an extended range of the current suppression, while an anti-ferromagnetic exchange leads to a decreased range of the blockade regime.
\end{abstract}
\maketitle

\section{Introduction}
\label{sec-intro}
The non-equilibrium properties of coupled quantum dots (QDs) in the Coulomb blockade regime, of both serial and parallel configurations,\cite{wiel2003,livermore1996,chan2002} have been studied extensively. Much of the interest in QD systems is of fundamental character because of the close analogy of QDs as artificial atoms and coupled QDs as artificial molecules.\cite{kastner1993} Especially, coupled QDs provide the opportunity to tune the interdot tunnelling and capacitive coupling which change the conductance spectra and current-voltage ($I-V$) characteristics of the system. QDs displaying ratchet effects have been explored both experimentally and theoretically,\cite{linke1998,linke1999} and it was recently shown that Coulomb interactions between QDs can give rise to ratchet effects.\cite{stopa2002,vidan2004} Other phenomena have been studied extensively in coupled QD systems through the past decade, such as, negative differential conductance,\cite{ishibashi2001,fransson12004,fransson22004} $I-V$ asymmetries,\cite{fransson22004,fransson32004} spin-pumping and spin-filtering,\cite{cota2005} quantum phase transitions,\cite{andrei1999} and the Kondo effect in serial\cite{georges1999,busser2000,aguado2000,jeong2001} and parallel\cite{aono1999,yshai2001,lopez2002} coupled QDs.

Recent transport measurements on double quantum dots (DQDs) have revealed technical possibilities to utilise the Pauli exclusion principle for the construction of efficient current rectifying devices.\cite{ono2002,johnson2005} The charge transport through a series of zero dimensional sites, with broken inversion symmetry with respect to the applied bias voltage, is suppressed by a combination of the Pauli principle and the Coulomb blockade in one direction while permitting it to flow in the opposite direction, thus, realising a controllable spin-Coulomb rectifier. In systems with absence of inversion symmetry, bias-dependent rectification effects can, on the one hand, be provided by bias voltage dependent transition rates between the states in the system, as was suggested for single molecules with appropriately configured molecular orbitals,\cite{aviram2000} and for multiple dots suffering from the Coulomb blockade effect.\cite{stopa2002} On the other hand, as we show in this paper, while the transition rates normally are bias voltage dependent, this is \emph{not} a necessary condition to achieve the spin-Coulomb, or \emph{Pauli spin}, blockade and rectifying properties of the system. Instead, for small interdot tunnelling rates we find it necessary that the two-electron triplet states are aligned (or almost aligned) with, at least, the lowest two-electron singlet state, and that these states are almost aligned with the lowest one-particle states. This condition for the DQD, described in more detail later in the paper, agrees very well with experimental set-up. While the system displays $I-V$ asymmetries and ratchet effects when this condition is not satisfied, the origin of those effects is more related to normal Coulomb blockade rather than the more intriguing Pauli spin blockade observed in Refs. \onlinecite{ono2002,johnson2005}.

In this paper we undertake a theoretical study of the transport properties of a DQD in the regime considered in Ref. \onlinecite{ono2002}, in order to obtain a qualitative picture of the origin of the Pauli spin blockade. In excellent agreement with the experimental report we find that the blockade regime indeed arises because the two-electron triplet state in the DQD is occupied with nearly a unit probability. The probability for the two-electron triplet state to be occupied strongly depends on the ratio between the interdot tunnelling rate $t$ and the energy spacing $\Delta\dote{}=\dote{A}-\dote{B}$, where $\dote{A/B}$ is the single electron level in QD$_{A/B}$. The DQD system itself provides a range of different regimes, of which the one considered here is given in the following set up, e.g. $|\Delta\dote{}|, U'\approx U/2$, and $\mu-\max\{\dote{A},\dote{B}\}\approx-U/2$, where $U'$ is the interdot Coulomb repulsion whereas $U$ is the on-site charging energy of QD$_{A/B}$, and $\mu$ is the equilibrium chemical potential. For definiteness, assume that the system is biased such that the electron flow is directed according to \emph{left lead} $\rightarrow$ QD$_{A}\ \rightarrow$ QD$_{B}\ \rightarrow$ \emph{right lead}. Then, for a small ratio $2t/\Delta\dote{}>0$, the probability for an electron in any of the two-particle singlet states to exit the DQD to the right lead is sufficiently large to prevent a build-up of the population in these states. The probability for electrons in the left lead to enter the DQD in any of the singlet states is of the same order which, however, does not affect the inability of charge accumulation in these states. For the two-electron triplet states the situation is somewhat different, since electrons occupying the triplet has a very low probability to exit the DQD to the right lead, whereas there is an almost unit probability for an electron in the left lead to enter the DQD in the two-electron triplet. Hence, there will be an accumulation of electron density in the triplet state, which leads to an almost complete occupation of this state. The accumulation of electron density in the two-electron triplet along with the negligible probability for occupying the two-electron singlets (as well as for the one- and three electrons states) provide a blockade of the current, e.g. Pauli spin blockade, for a finite range of bias voltages over the DQD. It is important to notice that the occupation probabilities are equally distributed among the spin $M_S=1,0,-1$ configurations. Therefore, although one can conclude that the Pauli spin blockade is caused by a fully occupied spin 1 state, it would in general be impossible to read out the definite spin state of the DQD in this regime. 

In the opposite case, e.g. small $2t/\Delta\dote{}<0$, the configuration of the DQD is mirrored in the sense that the triplet state acquires a small probability for occupation since the leakage into this state from the left lead is small whereas the corresponding leakage to the right lead is large. Thus, there cannot be an accumulation of charge in the triplet state. In contrast, since the probabilities for electrons to enter and exit the DQD via the singlet states are non-negligible (as in the former case), the occupation in the DQD oscillates between one and two electrons, that is, the current is not blockaded by a charge accumulation in the DQD.

The considered DQD system is complicated in the sense that we have included on-site charging energy of the QDs, the interdot Coulomb and hopping energies, as well as the interdot exchange interaction energy. It is therefore motivated to describe the physics of the DQD in terms of its eigenstate, given in the atomic limit (disconnected to the leads). The current through the DQD is thus described by considering single electron transitions between the various states. The treatment of the DQD dynamics is performed by means of a many-body density matrix approach for the population numbers of the eigenstates, up to the first order with respect to the couplings to the leads. This density matrix approach is similar to the one employed by Cota \etal\cite{cota2003} Also, Coish and Loss used this technique in their study of electron spin dynamics interacting with an environment of nuclear spins via hyperfine interactions.\cite{coish2004} In the model we only take one level in each QD, which is sufficient to describe the main features of the transport properties for low bias voltages, however, in the non-linear transport regime. For larger voltages than considered here, it is expected that more levels in the QDs participate in the conduction which would provide additional understanding to the experimental observations.\cite{ono2002} Our model differs from that used by Cota \etal,\cite{cota2003} in that we also include spin-spin interactions between electrons in the two QDs. Furthermore, while they considered time-dependent transport induced by spin-polarised pumping in the DQD, we concentrate here on the stationary Pauli spin blockade phenomenon which to our knowledge has not been considered before.

The rest of the paper begins with a definition and discussion of the model for the DQD in Sec. \ref{sec-model}, and the employed density matrix approach is described in Sec. \ref{sec-dm}. In Sec. \ref{sec-N} we analyse the equations for the population numbers, both analytically and numerically in a few different cases, while the paper is concluded in Sec. \ref{sec-conclusions}.

\section{Model of the system}
\label{sec-model}
In order to capture the main features of the DQD studied in Ref. \onlinecite{ono2002}, we consider two single-level quantum dots (QDs) with the intradot Coulomb energies $U_{A/B}$, interacting via the hopping interaction $t$, and interdot charging energy $U'$ and exchange $J$. Thus, we introduce the model, c.f. Refs. \onlinecite{sandalov1995,inoshita2003,cota2005},
\begin{eqnarray}
\Hamil_{DQD}&=&\sum_\sigma\dote{A\sigma}\ddagger{A\sigma}\dc{A\sigma}
	+U_An_{A\up}n_{A\down}
\nonumber\\&&
	+\sum_\sigma\dote{B\sigma}\ddagger{B\sigma}\dc{B\sigma}
	+U_Bn_{B\up}n_{B\down}
\nonumber\\&&\vphantom{\sum_\sigma}
	+(U'-J/2)(n_{A\up}+n_{A\down})(n_{B\up}+n_{B\down})
\nonumber\\&&
	-2J\bfs_A\cdot\bfs_B
	+\sum_\sigma t(\ddagger{A\sigma}\dc{B\sigma}+H.c.),
\label{eq-DQD}
\end{eqnarray}
where $\ddagger{A\sigma/B\sigma}\ (\dc{A\sigma/B\sigma})$ creates (annihilates) an electron in QD$_{A/B}$ at the energy $\dote{A\sigma/B\sigma}$ and spin projection $\sigma=\up,\down$, whereas $n_{A\sigma}=\ddagger{A\sigma}\dc{A\sigma}$ and likewise for $n_{B\sigma}$. The sixth term in Eq. (\ref{eq-DQD}) accounts for the spin-spin interactions between the QDs with $\bfs_A=(1/2)\sum_{\sigma\sigma'}\ddagger{A\sigma}\hat{\sigma}_{\sigma\sigma'}\dc{A\sigma'}$, and analogous for $\bfs_B$, where $\hat{\sigma}=(\sigma^x,\sigma^y,\sigma^z)$ is the vector of the Pauli spin matrices. Below we will only consider spin-degenerate QDs, hence, we set $\dote{A\sigma}=\dote{A}$ and $\dote{B\sigma}=\dote{B}$. This assumption is consistent with the experimental set-up of Ref. \onlinecite{ono2002} since there were no magnetic field applied or (any significant) spin-orbit splitting of the QD levels.

\begin{table}[b]
\caption{Schematic representation of the eigenstates for the model Hamiltonian $\Hamil_{DQD}$ in Eq. (\ref{eq-DQD}). Here, $\ket{\sigma}\equiv\ddagger{\sigma}\ket{0,1}$, where $\ket{0,1}=\ket{0}\ket{0}$ is the empty state of the DQD, whereas $\ket{\sigma}\ket{\sigma'}\equiv\ket{\sigma}_A\ket{\sigma'}_B=\ddagger{B\sigma'}\ddagger{A\sigma}\ket{0,1}$, and $\ket{\up\down}\ket{0}=\ddagger{A\down}\ddagger{A\up}\ket{0,1}$ and correspondingly for QD$_B$. The two-particle states $\ket{\Phi_{S=0}^{AB}}=[\ket{\up}\ket{\down}-\ket{\down}\ket{\up}]/\sqrt{2}$, $\ket{\Phi_{S=0}^A}=\ket{\up\down}\ket{0}$, and $\ket{\Phi_{S=0}^B}=\ket{0}\ket{\up\down}$. The coefficients $\alpha_n$, $\beta_n$, $A_n,\ B_n,\ C_n$, $\kappa_n$, and $\lambda_n$ depend on the internal parameters $\dote{A\sigma/B\sigma}$, $U$, $U'$, $J$ and $t$ of the DQD.}
\begin{center}
\begin{tabular}{l|l}
\hline\hline
$N=1$ &
$\ket{1,n}=
	\alpha_n\ket{\up}\ket{0}
	+\beta_n\ket{0}\ket{\up},\ n=1,3$
\\
&
$\ket{1,n}=
	\alpha_n\ket{\down}\ket{0}
	+\beta_n\ket{0}\ket{\down},\ n=2,4$
\\\hline
$N=2$ & 
$\ket{2,1}=\ket{\up}\ket{\up},\
	\ket{2,2}=\ket{\down}\ket{\down}$
\\
& $\ket{2,3}=[\ket{\up}\ket{\down}+\ket{\down}\ket{\up}]\sqrt{2}$
\\
& $\ket{2,n}=A_n\ket{\Phi_{S=0}^{AB}}+B_n\ket{\Phi_{S=0}^{A}}
	+C_n\ket{\Phi_{S=0}^{B}}$
\\
& $n=4,5,6$
\\\hline
$N=3$ & $\ket{3,n}=\kappa_n\ket{\up\down}\ket{\up}+\lambda_n\ket{\up}\ket{\up\down},\ n=1,3$
\\
& $\ket{3,n}=\kappa_n\ket{\up\down}\ket{\down}+\lambda_n\ket{\down}\ket{\up\down},\ n=2,4$
\\\hline
$N=4$ & 
$\ket{4,1}=\ket{\up\down}\ket{\up\down}$
\\\hline\hline
\end{tabular}
\end{center}
\label{tab-states}
\end{table}
Because of the inherent complexity of this model it is motivated to transform the Hamiltonian $\Hamil_{DQD}$ into diagonal form. Therefore, we need to find the eigenstates of the system, Eq. (\ref{eq-DQD}), which are given on the form $\ket{N,n}$, representing the $n$th state of the $N$-electron configuration, see Table \ref{tab-states}. Obviously, there are 16 different states of the DQD. The corresponding energies are denoted by $E_{01}=0,\ E_{1n},\ E_{2n},\ E_{3n},\ E_{41}$, respectively. Thus, the model given in Eq. (\ref{eq-DQD}) can be written in diagonal form according to
\begin{eqnarray}
\Hamil_{DQD}&=&\sum_{N=0}^4\sum_nE_{Nn}\h{n}{N}.
\label{eq-diagDQD}
\end{eqnarray}
Here, we have introduced the Hubbard operators\cite{hubbard11963,hubbard21963} $\X{nn'}{NN'}=\ket{N,n}\bra{N',n'}$ and the diagonal transitions $\h{n}{N}\equiv\X{nn}{NN}$. The diagonalisation process becomes clear by considering the expansion
\begin{eqnarray*}
\dc{A\sigma}&=&\sum_{NN'}\sum_{nn'}
	\ket{N,n}\bra{N,n}\dc{A\sigma}\ket{N',n'}\bra{N',n'}=
\nonumber\\&=&
	\sum_{N}\sum_{nn'}(\dc{A\sigma})^{nn'}_{NN+1}\X{nn'}{NN+1},
\end{eqnarray*}
where we have put $(\dc{A\sigma})^{nn'}_{NN+1}\equiv\bra{N,n}\dc{\sigma}\ket{N+1,n'}$, for a shorter notation. A similar transformation can be made for the operator $\dc{B\sigma}$. The summation runs over all states $\ket{N,n}$ and $\ket{N+1,n'}$ in the DQD, where we have used that $(\dc{A\sigma})^{nn'}_{NN'}=\delta_{N'N+1}(\dc{A\sigma})^{nn'}_{NN+1}$. The expansion above along with the algebra of the Hubbard operators, i.e. $\X{nn'}{NN'}\X{pp'}{MM'}=\delta_{N'M}\delta_{n'p}\X{np'}{NM'}$ yield the desired transformation of the Hamiltonian to the form in Eq. (\ref{eq-diagDQD}). It should be noticed that the transition matrix elements $\bra{N,n}\dc{A\sigma/B\sigma}\ket{N',n'}$ are crucial for the transport properties of the system, as will become clear later in the paper.

The total system, constituted of the DQD and the contacts to the left ($L$) and right $(R)$ of the DQD is given by
\begin{eqnarray}
\Hamil&=&\sum_{k\sigma\in L,R}\leade{k}\cdagger{k}\cc{k}
	+\Hamil_{DQD}
\nonumber\\&&
	+\sum_{k\sigma\in L}(v_{k\sigma}\cdagger{k}\dc{A\sigma}+H.c.)
\nonumber\\&&
	+\sum_{k\sigma\in R}(v_{k\sigma}\cdagger{k}\dc{B\sigma}+H.c.)
\label{eq-system}
\end{eqnarray}
where $\cdagger{k}\ (\cc{k}),\ k\sigma\in L/R,$ creates (annihilates) an electron in the left/right lead at the energy $\leade{k}$. The last two terms express the tunnelling between the contacts and the DQD. In order to obtain a uniform description of the physics in the system, it is desired to express these in terms of Hubbard operators as well. By the above observations, we note that the tunnelling can be written as
\begin{equation}
\Hamil_T=\sum_{k\sigma}[v_{k\sigma}\cdagger{k}
	\sum_{N}\sum_{nn'}(\dc{\sigma})^{nn'}_{NN+1}\X{nn'}{NN+1}+H.c.]
\label{eq-HT}
\end{equation}
where $\dc{\sigma}=\dc{A\sigma}$ or $\dc{\sigma}=\dc{B\sigma}$ depending on whether $k\sigma\in L$ or $k\sigma\in R$.
Note that the spin indices on $\cc{k}$ and $\dc{\sigma}$ have to coincide, since we neglect possible spin-flip transitions in the tunnel barriers. Such transitions, however, may occur within the DQD but will be neglected here.

\section{Density matrix for the many-body occupation numbers}
\label{sec-dm}
In the model we consider, there are 16 different many-body states, cf. Table \ref{tab-states}, and hence there are 16 occupation numbers ($P_{Nn},\ N=0,\ldots,4$) to calculate. These occupation numbers provide the probability for the state $\ket{N,n}$ to be populated. Here, this will be done by means of a density matrix for the many-body occupation numbers, e.g. $\rho=\{\X{nn'}{NN'}\}_{Nn,N'n'}$. However, since we are only interested in the evolution of the occupation numbers up to second order in the hybridisation $v_{k\sigma}$, we can neglect the influence from the off-diagonal components in the density matrix. We can safely neglect all the off-diagonal terms, in the second order approximation, since i) the off-diagonal terms only give contributions to the diagonal components of the fourth order and higher in $v_{k\sigma}$, and ii) the off-diagonal components are multiplied by off-diagonal transition matrix elements (to the fourth order) which in general are very small compared to the diagonal transition matrix elements.

In the present case we only consider stationary currents through the system, hence it is sufficient to investigate $\partial\rho/\partial t=0$ subject to the boundary condition $1=\sum_{Nn}P_{Nn}$. The equations take the form
\begin{eqnarray}
\ddt P_{01}&=&\frac{2}{\hbar}\im\sum_{k\sigma\gamma}v_{k\sigma}
	(\dc{\sigma})^{1n}_{01}\av{\cdagger{k}\X{1n}{01}},
\nonumber\\
\ddt P_{Nn}&=&
	-\frac{2}{\hbar}\im\sum_{k\sigma}\biggl(
		\sum_{n'}v_{k\sigma}(\dc{\sigma})^{n'n}_{N-1N}
			\av{\cdagger{k}\X{n'n}{N-1N}}
\nonumber\\&&
		-\sum_{n'}v_{k\sigma}(\dc{\sigma})^{nn'}_{NN+1}
			\av{\cdagger{k}\X{nn'}{NN+1}}
	\biggr),
\nonumber\\
\ddt P_{41}&=&
	-\frac{2}{\hbar}\im\sum_{k\sigma}
		\sum_n v_{k\sigma}(\dc{\sigma})^{n1}_{34}
			\av{\cdagger{k}\X{n1}{34}}.
\label{eq-dtN}
\end{eqnarray}
In the second equation above the electron number $N=1,2,3$. To the first order of approximation (in the Markovian limit), which is sufficient for describing the current in the second order approximation, the imaginary part of the correlation functions $\av{\cdagger{k}\X{nn'}{NN'}},\ N'=N\pm1$, are given by
\begin{eqnarray}
\im\av{\cdagger{k}\X{nn'}{NN'}}&=&-\pi v_{k\sigma}^*(\ddagger{\sigma})^{n'n}_{N'N}
	[f^+_\chi(\leade{k})P_{Nn}
\label{eq-corrfunc}\\&&
	-f^-_\chi(\leade{k})P_{N'n'}]\delta(\Delta_{N'n',Nn}-\leade{k}),
\nonumber
\end{eqnarray}
where $f^+_\chi(\leade{k})=f(\leade{k}-\mu_\chi),\ k\sigma\in\chi=L,R$, is the Fermi function whereas $f^-_\chi(\leade{k})=1-f^+_\chi(\leade{k})$, and $\Delta_{N'n',Nn}=E_{N'n'}-E_{Nn}$ denote the energy for the transition $\X{nn'}{NN'}$. Introducing the couplings to the left and right leads by $\Gamma^{L/R}_{Nn,N'n'}=2\pi\sum_{k\sigma\in L/R}|v_{k\sigma}(\dc{\sigma})^{nn'}_{NN'}|^2\delta(\Delta_{N'n',Nn}-\leade{k})$, we find that the equations for the many-body population numbers can be written as
\begin{widetext}
\begin{subequations}
\label{eq-dtN}
\begin{eqnarray}
\ddt P_{01}&=&-\frac{1}{\hbar}\sum_{\chi n}\Gamma^\chi_{01,1n}
	[f^+_\chi(\Delta_{1n,01})P_{01}
	-f^-_\chi(\Delta_{1n,01})P_{1n}]
\label{eq-01}\\
\ddt P_{Nn}&=&\frac{1}{\hbar}\sum_\chi\biggl(
	\sum_{n'}\Gamma_{N-1n',Nn}^\chi
		[f^+_\chi(\Delta_{Nn,N-1n'})P_{N-1n'}
			-f^-_\chi(\Delta_{Nn,N-1n'})P_{Nn}]
\nonumber\\&&
	-\sum_{n'}\Gamma^\chi_{Nn,N+1n'}
		[f^+_\chi(\Delta_{N+1n',Nn})P_{Nn}
			-f^-_\chi(\Delta_{N+1n',Nn})P_{N+1n'}]\biggr)
\label{eq-Nn}\\
\ddt P_{41}&=&\frac{1}{\hbar}\sum_{\chi n}
	\Gamma_{3n,41}^\chi
		[f^+_\chi(\Delta_{41,3n})P_{3n}
			-f^-_\chi(\Delta_{41,3n})P_{41}].
\label{eq-41}
\end{eqnarray}
\end{subequations}
\end{widetext}

\section{Analysis of the population numbers}
\label{sec-N}
\begin{figure}[b]
\begin{center}
\includegraphics[width=8.5cm]{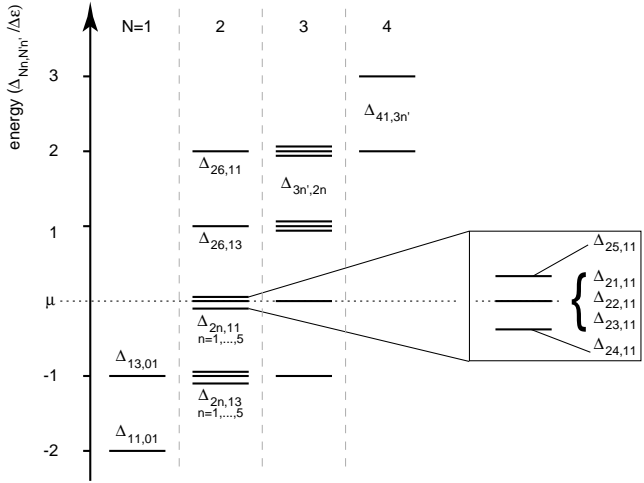}
\end{center}
\caption{Equilibrium distribution of the one-electron transition energies $\Delta_{Nn,N'n'}$ in terms of $\Delta\dote{}$, for $\dote{A}=-U/2$, $\Delta\dote{}=\dote{A}-\dote{B}=U/2$, $U'=U/2$, $2t/\Delta\dote{}\ll1$, and $J=0$. In the absence of magnetic field we have $\Delta_{12,01}=\Delta_{11,01}$, $\Delta_{14,01}=\Delta_{13,01}$, $\Delta_{26,12}=\Delta_{26,11}$, and $\Delta_{26,14}=\Delta_{26,13}$.}
\label{fig-states}
\end{figure}
The main task of this paper is to show that the Pauli spin blockade is caused by a full occupation of the two-electron triplet in the DQD. For this purpose, we tune the system into the conditions that were established in the experimental situation of Ref. \onlinecite{ono2002}, with two identical QDs. There, the internal parameters of the DQD were configured such that the levels of the two QDs were separated by $\Delta\dote{}=\dote{A}-\dote{B}=U/2$, where $U=U_A=U_B$, and that $\dote{A}-\mu=-U/2$. Moreover, the interdot Coulomb repulsion $U'=U/2$ and the interdot hopping $2t/\Delta\dote{}\ll1$. We note that the exchange interaction parameter $J$ in the experiment was estimated to be much less than any other parameter in the system, e.g. $J\ll k_BT\sim U/10$, thus we set $J=0$. The one-electron transition energies between the different states are graphically displayed in Fig. \ref{fig-states}. Under those circumstances we find that only $|\Delta_{2n,1n'}-\mu|\ll U/2$ for $n=1,\ldots,5$, and $n'=1,2$, and $|\Delta_{3n,26}-\mu|\ll U/2$ for $n=3,4$, where $\mu$ is the equilibrium chemical potential of the system. For all other transitions we have $|\Delta_{Nn,N-1n'}-\mu|\gtrsim U/2$, c.f. Fig \ref{fig-states}.

\subsection{The Pauli spin blockade regime}
\label{sec-spinC}
\begin{figure}[t]
\begin{center}
\includegraphics[width=6cm]{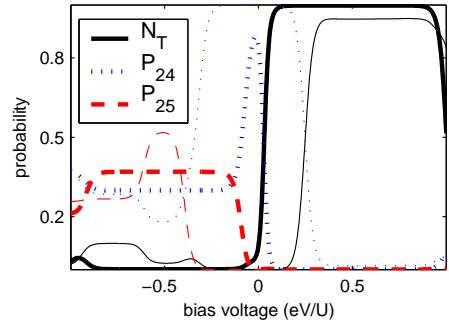}
\end{center}
\caption{(Colour online) The two-electron population numbers $N_T$ (solid), $P_{24}$ (dotted), and $P_{25}$ (dashed), for $\xi=0.1$ (bold) and $\xi=0.5$ (faint), holding $\Delta\dote{}$ constant, and $k_BT\sim U/10$. Other parameters as in Fig. \ref{fig-states}.}
\label{fig-Nt}
\end{figure}

Now, we let the bias voltage\cite{biasvoltage} $eV=\mu_L-\mu_R\in[0.1,1]U$ such that $f^+_L(\Delta_{2n,1n'})=1,\ f^+_R(\Delta_{2n,1n'})=0,\ n=1,\ldots,5,\ n'=1,2$. In this regime, we also note that $f^+_\chi(\Delta_{1n',01})=1$ for all $n'$, $f^+_\chi(\Delta_{2n,1n'})=1$, $n=1,\ldots,5,\ n'=3,4$, $f^+_\chi(\Delta_{26,1n'})=0$ for all $n'$, $f^+_\chi(\Delta_{3n',2n})=0$, $n'=1,\dots,4,\ n=1,\ldots,5$, $f^+_\chi(\Delta_{3n',26})=1$, $n'=1,2$, $f^+_L(\Delta_{3n',26})=f^-_R(\Delta_{3n',26})=1$, $n'=3,4$, and that $f^+_\chi(\Delta_{41,3n'})=0$ for all $n'$. Then, from Eq. (\ref{eq-01}) and (\ref{eq-41}) it follows that $P_{01}=P_{41}=0$, as expected since the transition energies $\Delta_{1n',01}$ and $\Delta_{41,3n'}$ lie far from resonance. It also follows that $P_{26}$ vanishes, which is clear from the fact that the transition energies $\Delta_{26,1n'}-\mu\gtrsim U/2$. Therefore, there cannot occur any transitions between the state $\ket{2,6}$ and any of the three-electron states, although the transition energies $\mu_R<\Delta_{3n',26}<\mu_L,\ n=3,4$. Taking this observation together with the fact that $\Delta_{3n',2n}-\mu\gtrsim U/2$, $n=1,\ldots,5,\ n'=1,\ldots,4$, it is clear that the population numbers $P_{3n'}=0$ for all $n'$. This implies that the population numbers $P_{1n'}$ of the one-electron states $\ket{1,n'}$, $n'=3,4$ vanish. Hence, the only non-vanishing population numbers are $P_{1n'},\ n'=1,2$ and $P_{2n},\ n=1,\ldots,5$, which lead to the equations
\begin{table}[b]
\caption{Matrix elements for the transitions $\X{n'n}{12}$ given in terms of the parameters of the eigenstates of the model, Eq. (\ref{eq-DQD}), c.f. Table \ref{tab-states}. Here $\alpha=\alpha_n,\ \beta=\beta_n,\ n=1,2$.}
\label{tab-me}
\begin{center}
\begin{tabular}{l}
\hline
$|(\dc{A\up})^{11}_{12}|^2=|(\dc{A\down})^{22}_{12}|^2=\beta^2$\\
$|(\dc{A\up})^{12}_{12}|^2=|(\dc{A\down})^{21}_{12}|^2=0$\\ $|(\dc{A\up})^{13}_{12}|^2=|(\dc{A\down})^{23}_{12}|^2=\beta^2/2$\\ $|(\dc{A\up})^{1n}_{12}|^2=|(\dc{A\down})^{2_n}_{12}|^2
	=(\beta A_n/\sqrt{2}+\alpha B_n)^2,\ n=4,5$\\
$|(\dc{B\up})^{11}_{12}|^2=|(\dc{B\down})^{22}_{12}|^2=\alpha^2$\\ $|(\dc{B\up})^{12}_{12}|^2=|(\dc{B\down})^{21}_{12}|^2=0$\\ $|(\dc{B\up})^{13}_{12}|^2=|(\dc{B\down})^{23}_{12}|^2=\alpha^2/2$\\ $|(\dc{B\up})^{1n}_{12}|^2=|(\dc{B\down})^{2n}_{12}|^2
	=(\alpha A_n/\sqrt{2}+\beta C_n)^2,\ n=4,5$\\
\hline
\end{tabular}
\end{center}
\end{table}
\begin{subequations}
\label{eq-None}
\begin{eqnarray}
P_{1n'}&=&\frac{\sum_{n=1}^5\Gamma_{1n',2n}^RP_{2n}}
	{\sum_{m=1}^5\Gamma_{1n',2m}^L},\ n'=1,2,
\label{eq-P1n}\\
0&=&\sum_{n'=1,2}[\Gamma_{1n',2n}^LP_{1n'}
	-\Gamma_{1n',2n}^RP_{2n}],
\nonumber\\&&
	n=1,\ldots,5,
\label{eq-P2n}
\end{eqnarray}
\end{subequations}
where we have used that $f^+_L(\Delta_{2n,1n'})=1$ and $f^+_R(\Delta_{2n,1n'})=0$. The three configurations of the two-electron triplet, e.g. $\ket{2,n},\ n=1,2,3$, have to have equal probability (spin degenerate system), hence, we set $P_{2n}=N_T/3,\ n=1,2,3$. The transition matrix elements $(\dc{\sigma})^{n'n}_{12}$ are in general given in terms of the internal parameters of the DQD, see Table \ref{tab-me}. Using these we find that $P_{11}=P_{12}=N_1/2$, which is expected since the DQD is spin degenerate. From Eq. (\ref{eq-None}) we then also find that
\begin{subequations}
\label{eq-NasNT}
\begin{eqnarray}
N_1&=&\frac{2}{3}\frac{\Gamma^R}{\Gamma^L}
	\biggl(\frac{\alpha}{\beta}\biggr)^2N_T
\label{eq-P1nT}\\
P_{2n}&=&\frac{1}{3}
	\biggl(
		\frac{L_n}{R_n}
		\cdot\frac{\alpha}{\beta}\biggr)^2N_T,\ n=4,5,
\label{eq-P2nT}
\end{eqnarray}
\end{subequations}
where $L_n=\beta A_n/\sqrt{2}+\alpha B_n$ and $R_n=\alpha A_n/\sqrt{2}+\beta C_n$, which are bounded and finite for all $\xi=2t/\Delta\dote{}$ and slowly varying functions of $\xi$. Thus, by the normalisation condition, now reduced to $1=N_1+N_T+P_{24}+P_{25}$, we finally arrive at
\begin{equation}
N_T=\biggl\{1+\frac{1}{3}\biggl(\frac{\alpha}{\beta}\biggr)^2
	\biggl[2\frac{\Gamma^R}{\Gamma^L}+\biggl(\frac{L_4}{R_4}\biggr)^2
		+\biggl(\frac{L_5}{R_5}\biggr)^2\biggr]
		\biggr\}^{-1}.
\label{eq-NT}
\end{equation}
Hence, the probability for populating the two-electron triplet in the restricted bias voltage interval can be given in terms of the internal parameters of the DQD and the couplings to the leads. Especially, we observe that $N_T$ approaches 1 for $\xi\ll1$, since $\alpha^2=\xi^2/[(1+\sqrt{1+\xi^2})^2+\xi^2]\approx0$ and $\beta^2=(1+\sqrt{1+\xi^2})^2/[(1+\sqrt{1+\xi^2})^2+\xi^2)\approx1$. This is consistent with Eq. (\ref{eq-NasNT}), which shows that $N_1$ and $P_{2n},\ n=4,5$, tend to zero in the same limit. The main conclusion then is that the occupation of the two-electron triplet strongly depends on the ratio between the interdot hopping strength $t$ and the relative level separation  $\Delta\dote{}$, e.g. $\xi$, and that the probability of a fully occupied two-electron triplet approaches unity as $\xi\rightarrow0$. One should note, however, that this conclusion follows independently of a possible bias dependence of the transition matrix elements. Including such a dependence into the consideration only strengthens our conclusion. A similar analysis can be performed with a finite exchange interaction parameter $J$, which will result in an equivalent conclusion of the existence of a Pauli spin blockade regime, see Sec. \ref{sec-J}. In Fig. \ref{fig-Nt} we display the bias dependence of the two-electron population numbers $N_T,\ P_{2n},\ n=4,5$, for two different values of $\xi$, and it is clear that the two-electron triplet state is almost fully populated in the considered regime of bias voltages for small $\xi$, whereas it becomes depopulated for increasing $\xi$. It is also clear from the figure that $P_{2n},\ n=4,5$, are negligible in the spin-Coulomb regime.

The current $I$ through the DQD will be suppressed by the high population of the two-electron triplet state. In the stationary regime $I=(I_L-I_R)/2,\ I_L=-I_R$, where the current $I_\chi$ flowing from the $\chi$ lead to the DQD is given by
\begin{equation}
I_\chi=-\frac{e}{h}\tr\im\int\bfGamma^\chi[f^+_\chi(\omega)\bfG^>(\omega)
	+f^-_\chi(\omega)\bfG^<(\omega)]d\omega,
\label{eq-JL}
\end{equation}
where the greater and lesser Green functions of the DQD in the zeroth order approximation are given by\cite{fransson12004,fransson22004,fransson2005} $G_{Nn,N+1n'}^<(\omega)=i2\pi P_{N+1n'}\delta(\omega-\Delta_{N+1n',Nn})$ and $G_{Nn,N+1n'}^>(\omega)=-i2\pi P_{Nn}\delta(\omega-\Delta_{N+1n',Nn})$, respectively. Here, the coupling matrix $\bfGamma^\alpha$ includes the transition matrix elements of the DQD. Using the above analysis for the population numbers we find that the current $I_L^\rightarrow$ in the Pauli spin blockade regime reduces to
\begin{eqnarray}
I_L^\rightarrow&=&\frac{2e}{3\hbar}\Gamma^R\biggl(\frac{\alpha}{\beta}\biggr)^2
	[3\beta^2/2+L_4^2+L_5^2]N_T
\label{eq-JV+}\\
&=&\frac{2\pi}{3}\frac{e\Gamma}{h}\biggl(\frac{\alpha}{\beta}\biggr)^2
	[3\beta^2/2+L_4^2+L_5^2]N_T,
\nonumber
\end{eqnarray}
where the last line is obtained in the case of symmetric couplings, $\Gamma^L=\Gamma^R=\Gamma/2$. As expected, the current is proportional to the population number $N_T$ of the two-electron triplet state. More importantly, though, is that the current strongly depends on the ratio $(\alpha/\beta)^2=\xi^2/(1+\sqrt{1+\xi^2})^2$, showing a strong suppression of the current for weakly coupled QDs ($\xi\ll1$).

\begin{figure}[t]
\begin{center}
\includegraphics[width=6cm]{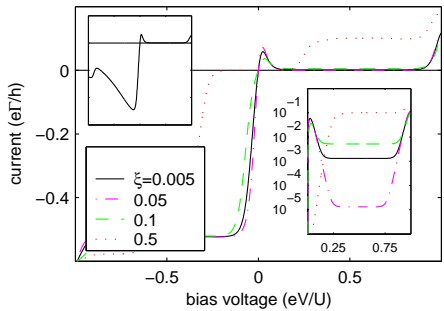}
\end{center}
\caption{(Colour online) The $I-V$ characteristics of the DQD system for various values of $\xi$, holding $\Delta\dote{}$ constant, $k_BT\sim U/10$, and $\Gamma=2\Gamma^L$. The lower right inset shows the current in the Pauli spin blockade regime. The upper left inset shows the current for bias voltage dependent transition matrix elements imposing a 10 \% voltage drop between the QDs, and $\xi=0.05$ (same scale as main panel).}
\label{fig-JVt}
\end{figure}
In Fig. \ref{fig-JVt} we show the $I-V$ characteristics of the DQD system in the given configuration for different $\xi$, holding $\Delta\dote{}$ constant. This shows a clearly suppressed current in the forward biased system, e.g. $eV=\mu_L-\mu_R>0$. The lower right inset shows the forward current in the Pauli spin blockade regime on a logarithmic scale, and for small $\xi$ it is readily seen that the current is about 1 | 4 orders of magnitude smaller in the blockaded regime than in the linear regime around zero voltage. This verifies the experimental observation presented in Ref. \onlinecite{ono2002}, since bias voltage dependent transition matrix elements merely provide a slight modification of the current, see upper left inset of Fig. \ref{fig-JVt} where we have followed the experimental estimate of a 10 \% voltage drop between the QDs. The bias dependence of the transition matrix elements is provided by assuming a bias dependent shift of the single particle levels in the QDs.

The current $I_L^\rightarrow$ is non-vanishing in the spin-Coulomb regime because of the finite hopping interaction $t$ between the QDs, which causes the triplet state population number $N_T$ to be slightly less than unity. The remainder of the available population is distributed among the one-particle states $\ket{1,n},\ n=1,2$, and the two-electron singlets $\ket{2,n},\ n=4,5$. The non-vanishing population number $N_1$ implies that there is a finite probability for one electron to exit the DQD to the right lead, and thereafter another electron enters the DQD from the left. The leakage current in the blockaded regime grows quadratically with $\xi$ since $(\alpha/\beta)^2=\xi^2/(1+\sqrt{1+\xi^2})^2$, while $N_T$ decays by $1/(1+(\alpha/\beta)^2/3)$, see Figs. \ref{fig-Nt} and \ref{fig-JVt}.

\begin{figure}[t]
\begin{center}
\includegraphics[width=6cm]{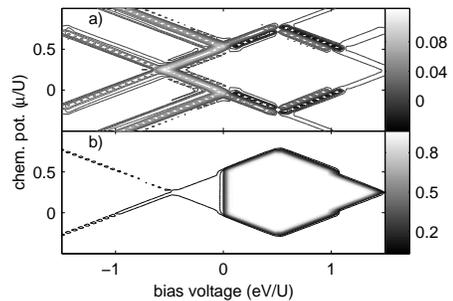}
\end{center}
\caption{a) Differential conductance $(dI/dV)$ of the DQD for varying positions of the chemical potential $\mu\in[-1,2]U$. The areas between the grey (black) ridges correspond to $dI/dV=0$. b) Contour plot of the population number $N_T=1$ as function of the bias voltage and the position of the chemical potential. The truncated diamond in the bias voltage range $[0,1.5]U$ marks the region with $N_T=1$. Here, $\xi=0.05$ while other parameters as in Fig. \ref{fig-Nt}.}
\label{fig-NTsurf}
\end{figure}
In the above analysis of the population numbers, the equilibrium chemical potential $\mu=0$. Shifting the position of $\mu$ corresponds to applying a gate voltage over the system, which changes the positions of the QD levels relatively $\mu$. In doing this we find that the Pauli spin blockade regime extends over a range of gate voltages, where $N_T\approx1$ and the current through the DQD is small. In Fig. \ref{fig-NTsurf} b) we show a contour plot of $N_T$ as function of the bias voltage and $\mu$, readily showing the expected domain of the Pauli spin blockade regime (truncated diamond in the bias voltage range $(0,1.5)U$). The corresponding differential conductance of the DQD system is plotted in Fig. \ref{fig-NTsurf} a), where the grey and white ridges correspond to $dI/dV>0$ while the black ridges corresponds to $dI/dV<0$. The areas between the ridges correspond to vanishing differential conductance.

\begin{figure}[t]
\begin{center}
\includegraphics[width=6cm]{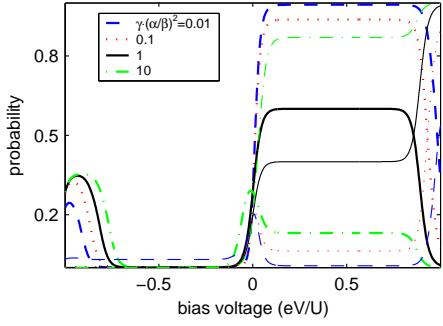}
\end{center}
\caption{(Colour online) The two-electron population number $N_T$ (bold) for different asymmetries $\gamma=\Gamma^R/\Gamma^L$ of the couplings to the leads. The sum of the one-electron population numbers $N_1$ (faint) are plotted for reference. Here, $\xi=0.05$, whereas other parameters as in Fig. \ref{fig-Nt}.}
\label{fig-Ngamma}
\end{figure}
The population number $N_T$ of the two-electron triplet state, which strongly depends on $(\alpha/\beta)^2$, depends on the couplings $\Gamma^{L/R}$ to the leads as $\Gamma^R/\Gamma^L$, see Eq. (\ref{eq-NT}). Hence, assuming proportional couplings, i.e. $\Gamma^R/\Gamma^L=\gamma\geq0$, such that $\gamma\sim(\beta/\alpha)^2$, provides a lifting of the Pauli spin blockade even for weakly coupled QDs. Accordingly, for $\xi\rightarrow0$ one finds that $N_T\rightarrow3/5$. In Fig. \ref{fig-Ngamma} we plot the population number $N_T$ (bold) for different asymmetries of the couplings to the left and right leads, and it is readily seen that $N_T$ decreases as $\gamma$ is increased. Especially, for $\gamma=(\beta/\alpha)^2$ it is seen that $N_T$ is close to 3/5. The population numbers $P_{2n},\ n=4,5$, remain negligible as $\gamma$ is increased since they do not depend on this ratio. The remaining population $(1-N_T)$ is distributed among the one-electron states $\ket{1n},\ n=1,2$, see Eq. (\ref{eq-P1nT}), since $N_1\propto\gamma N_T$. In Fig. \ref{fig-Ngamma} we plot the population number $N_1$ (faint) of the one-electron states, illustrating that the lifted Pauli spin blockade results in an increasing population in the one-electron states.

The absolute magnitude of the current, however, is not increased by the lifting of the Pauli spin blockade. In the limit $\gamma=x(\beta/\alpha)^2$ and $\xi\ll1$, where $x>0$ is an arbitrary constant, the current becomes (using $\Gamma=\Gamma^L+\Gamma^R=(1+\gamma)\Gamma^L$)
\begin{equation}
I_L^\rightarrow\approx\frac{4xe}{(3+2x)\hbar}\Gamma^L
	=\frac{8\pi x}{3+2x}\frac{1}{1+\gamma}	\frac{e\Gamma}{h},
\label{eq-JVlifted}
\end{equation}
showing that the current remains small (decreasing) for increasing asymmetries $\gamma$ of the couplings to the leads. Nevertheless, as we will see in the next paragraph, Sec. \ref{sec-rbias}, the current $I_L^{\leftarrow}$ in the reverse bias direction (in the range $-[1,0.1]U$) becomes in this limit $I_L^{\leftarrow}\approx[\pi/(1+\gamma)]\cdot e\Gamma/h$. It is, thus, clear that the ratio between the forward and backward currents, e.g. $I_L^\rightarrow/I_L^\leftarrow$ is a bounded and monotonically increasing function of the parameter $\gamma$. Hence, although the absolute magnitude of the current is not increased by the lifting of the Pauli spin blockade caused by the asymmetric couplings to the leads, the forward current increases relatively the backward current for growing asymmetries $\Gamma^R/\Gamma^L$.

\subsection{Reverse bias}
\label{sec-rbias}
An analogous investigation of the case $eV=\mu_L-\mu_R\in-[1,0.1]U$ shows that the population number $N_T$ of the two-electron triplet approaches zero for $\xi\ll1$. This is easiest seen by expressing the population numbers $N_1,\ N_T,\ P_{25}$ in terms of $P_{24}$, which give
\begin{subequations}
\label{eq-Nas24}
\begin{eqnarray}
N_1&=&2\frac{\Gamma^L}{\Gamma^R}\biggl(\frac{L_4}{R_4}\biggr)^2P_{24}
\label{eq-N124}\\
N_T&=&\frac{3}{2}\biggl(\frac{\alpha}{\beta}\cdot\frac{L_4}{R_4}\biggr)^2P_{24},
\label{eq-NT24}\\
P_{25}&=&\biggl(\frac{L_4}{R_4}\cdot\frac{R_5}{L_5}\biggr)^2P_{24}.
\label{eq-P2424}
\end{eqnarray}
\end{subequations}
These equations together with the normalisation yields
\begin{equation}
P_{24}=\biggl\{1+\biggl(\frac{L_4}{R_4}\biggr)^2\biggl[2\frac{\Gamma^L}{\Gamma^R}
	+\frac{3}{2}\biggl(\frac{\alpha}{\beta}\biggr)^2
	+\biggl(\frac{R_5}{L_5}\biggr)^2\biggr]\biggr\}^{-1}.
\label{eq-P24}
\end{equation}
Thus, for small $\xi$ it is readily seen that $N_T\approx0$. This also shows that the population of the one- and two-electron states $\ket{1,n'},\ n'=1,2$ and $\ket{2,n},\ n=4,5$, respectively, are finite, meaning that the electron population in the DQD varies between 1 and 2. The reverse bias case corresponds to negative biases in Fig. \ref{fig-Nt}, showing the small triplet state population number $N_T$ and finite singlet population numbers $P_{24}$ (dotted) and $P_{25}$ (dashed). In the considered regime, the sum $P_{24}+P_{25}<1$, and since $N_T$ is negligible the remaining population is distributed among the one-electron states, e.g. population number $N_1$. Hence, the electron population in the DQD flips between 1 and 2 which allows a finite current to run through the system. The current $I_L^\leftarrow$ in the reverse direction, given by
\begin{eqnarray}
I_L^\leftarrow&=&\frac{2e}{\hbar}\Gamma^LL_4^2[1+(R_5/R_4)^2]P_{24}.
\label{eq-JL-}\\
&=&2\pi\frac{e\Gamma}{h}L_4^2[1+(R_5/R_4)^2]P_{24}.
\end{eqnarray}
is expected to be substantially larger than in the forward direction. Here, the last line is again obtained in the case of symmetric coupling.

\subsection{Linear regime}
\label{sec-linear}
In the linear regime, e.g. $eV\in(-0.1,0.1)U$, we can employ the equilibrium result
\begin{subequations}
\label{eq-Neq24}
\begin{eqnarray}
N_1&=&2e^{(\Delta_{24,11}-\mu)/k_BT}P_{24}
\label{eq-N1eq24}\\
N_T&=&3e^{(\Delta_{24,11}-\Delta_{21,11})/k_BT}P_{24}
\label{eq-NTeq24}\\
P_{25}&=&e^{(\Delta_{24,11}-\Delta_{25,11})/k_BT}P_{24},
\label{eq-P25eq24}
\end{eqnarray}
\end{subequations}
which gives
\begin{eqnarray}
P_{24}&=&\{1+2e^{(\Delta_{24,11}-\mu)/k_BT}+3e^{(\Delta_{24,11}-\Delta_{21,11})/k_BT}
\nonumber\\&&
	+e^{(\Delta_{24,11}-\Delta_{25,11})/k_BT}\}^{-1}.
\label{eq-P24eq}
\end{eqnarray}
Since $\Delta_{24,11}<\mu=0$, see Fig. \ref{fig-states}, we see from these relations that $P_{24}\approx1$ whenever $|\Delta_{24,11}|\gg k_BT$ which is satisfied for $\xi\gtrsim0.1$ and $k_BT\sim U/10$. Accordingly, the population numbers in Eq. (\ref{eq-Neq24}) will be exponentially suppressed by the same condition on $\xi$, since the gap between the energy of the lowest two-electron singlet, e.g. $\ket{2,4}$, and the other two-electron energies increase roughly by $\xi^2$. Hence, for $\xi\gtrsim0.1$ the current will be blockaded by the lowest two-electron singlet in a finite bias voltage interval around equilibrium. The plots in Fig. \ref{fig-Nt} and \ref{fig-JVt} illustrates the blockade because of the fully occupied singlet state $P_{24}$ in the region around zero bias voltage for large $\xi$. 
It is also clear that $P_{24}\rightarrow1/7$ as $\xi\rightarrow0$, since the energy distance between $\Delta_{24,11}$ and $\mu,\ \Delta_{2n,11},\ n=1,2,3,5$, approach zero. In the limit of weakly coupled QDs, e.g. $\xi\rightarrow0$, the equilibrium population numbers satisfy $N_1=2/7,\ N_T=3/7,\ P_{24}=P_{25}=1/7$. The conductance of the system in the linear regime for weakly coupled QDs is therefore substantially larger than in the Pauli spin blockade regime.

\subsection{The role of the interdot exchange}
\label{sec-J}
In the previous sections we have considered the exchange interaction parameter $|J|\ll k_BT$, in which case the singlet-triplet splitting is negligible. Here, we will consider the case $|J|>2k_BT$, where $J>0$ gives a ferromagnetic interdot exchange in our model whereas $J<0$ provides an anti-ferromagnetic interdot exchange. The energy of the triplet state is given by $E_{2n}=\dote{A}+\dote{B}+U'-J$, $n=1,2,3$, whereas the lowest singlet state energy is roughly $E_{\check{S}}\equiv\min_{n=4,5,6}\{E_{S2n}\}\approx\dote{A}+\dote{B}+U'+J-F(t,\Delta\dote{})$, where $F(t,\Delta\dote{})$ is a positive function. For weakly coupled QDs, e.g. $\xi\ll1$, we find that $F(t,\Delta\dote{})\approx\xi$. Hence, in the configuration of the DQD system of interest in this paper and for low temperatures, the lowest singlet state is the ground state of the DQD whenever $J<\xi/2$ while the triplet becomes the ground state for larger $J$.

Here, the task is to demonstrate, for $\xi<2k_BT$, that there is a finite range of bias voltages population probability of the triplet state is almost unity, at least for interdot exchange interactions $-0.5(U-U')/2<J<(U-U')/2$. Indeed, when the system is biased such that $\mu_R<\Delta_{2n,1n'}<\mu_L$, $n=1,\ldots,5,\ n'=1,2$, while all other transition energies lie above or below both $\mu_L$ and $\mu_R$, we can apply the same arguments as in Sec. \ref{sec-spinC}. Then, it follows that $N_T\rightarrow1$ whenever $\xi\rightarrow0$. This argument is valid for all $2J/(U-U')\in(-0.5,1)$. It is clear, however, that this range of interdot exchange interactions contains more interesting features.

First, consider the range $0\leq J\lesssim\xi$. While the lowest singlet state remains the ground state of the DQD in this regime, the transport properties of the system are not significantly changed from the case of vanishing $J$. Hence, the Pauli spin blockade regime is again found for positive biases, however, slightly extended with larger $J$ since the transition energies $\Delta_{2n,1n'}-\mu<0$, $n=1,2,3,\ n'=1,2$.

Secondly, for $J$ well above $\xi$, the triplet state acquires the lowest energy among the two-electron states, e.g. $\Delta_{2n,1n'}-\min_{m=4,5,6}\{\Delta_{2m,1n'}\}$, $n=1,2,3,\ n'=1,2$. Hence, depending on the value of $J$ in the interval $(\xi,[U-U']/2)$, the Pauli spin blockade regime is extended to include also equilibrium as well as a finite interval of negative biases. The reason for this is to be found in the matrix elements of the transitions between the one-electron states and the lowest singlet states, while the matrix elements for transitions between the one-electron states and the triplet are independent of $J$. For small $J$, the two lowest singlet states are almost equally weighted on the Fock states $\ket{\Phi^{AB}_{S=0}}$ and $\ket{\Phi^B_{S=0}}$ (see Table \ref{tab-states} for the definitions of these states). Hence, those state are coupled to the left and right lead with almost equal strength. For large $J$, however, the weights of the singlets are redistributed among the Fock states such that the lowest singlet acquires a larger weight on QD$_B$, e.g. on the state $\ket{\Phi^B_{S=0}}$, whereas the other is more strongly weighted on $\ket{\Phi^{AB}_{S=0}}$. This, in turn, leads to that electrons tunnelling to the lowest singlet state in the DQD from the left/right lead is small/large which prevents accumulation of electron density in this state for biases such that $\Delta_{2n,1n'}-\mu_\chi<0$, $n=1,2,3,\ n'=1,2$, and $\min_{\chi=L,R}\{\mu_\chi\}\leq\min_{m=4,5,6}\{\Delta_{2m,1n'}\}\leq\max_{\chi=L,R}\{\mu_\chi\}$. For larger negative biases, e.g. such that $\mu_L<\Delta_{2n,1n'}<\mu_R$, $n=1,2,3,\ n'=1,2$, the population in the triplet is lifted and the current can flow through the DQD via transitions between the one-electron states and the lowest singlet.

Third, a anti-ferromagnetic exchange, e.g. $J<0$, yields a smaller range of positive bias voltages (compared to the case $J=0$) where the triplet state is fully populated. This is understood as an effect of that $\Delta_{2n,1n'}-\mu>0$, $n=1,2,3,\ n'=1,2$. Hence, a larger bias voltage is required (than in the case of $J=0$) to enter the Pauli spin blockade regime.

\begin{figure}[t]
\begin{center}
\includegraphics[width=6cm]{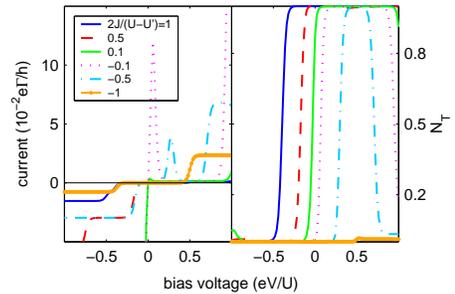}
\end{center}
\caption{(Colour online) $I-V$ characteristics a) and triplet state population probability b) for various strengths of the interdot exchange interaction. Here $\xi=0.05$ and other parameters as in Fig. \ref{fig-Nt}.}
\label{fig-JNTJ}
\end{figure}
In Fig. \ref{fig-JNTJ} we display the $I-V$ characteristics a) and triplet state population probability for several values of the interdot exchange interaction parameter $J$. The plots in the figure clearly shows that the transport properties of the system for $|2J/(U-U')|<0.1$ resembles the case of negligible $J$. It is however, also clear from these plots that the Pauli spin blockade regime is entered for lower biases as $J>0$ and for higher biases as $J<0$, as expected from the above discussion. The case $2J/(U-U')=-1$ has been included for completeness, where the blockade around equilibrium is caused by a full occupation in the lowest singlet state, that is, conventional Coulomb blockade.

\section{Conclusions}
\label{sec-conclusions}
We have theoretically verified the experimentally observed spin-Coulomb, or Pauli spin, blockade\cite{ono2002} caused by a full occupation of the two-electron triplet state in a DQD. Tuning the system such that the level off-set $\Delta\dote{}=\dote{A}-\dote{B}=\mu-\dote{A}$ and the interdot Coulomb repulsion are half of the on-site charging energy $U$, provide a unit probability for populating the two-electron triplet state in one direction of the bias voltage for small interdot tunnelling $t$. The almost full population of the two-electron triplet state prevents a current flow through the DQD, however, a small leakage current flows through the structure because of non-vanishing populations in the one-electron states and the lowest two-electron singlet state. Depending on the ratio $2t/\Delta\dote{}$, the leakage current in the blockaded regime is around 1 | 4 orders of magnitude smaller than the current in the near outside of this regime. The population of the triplet state is lifted in the other bias direction which allows a substantial increase of the current through the system. We have also shown that a finite ferromagnetic exchange interaction between the QDs extends the range of the Pauli spin blockade regime, whereas an anti-ferromagnetic exchange yields a smaller range of the Pauli spin blockade regime.

\acknowledgments
J. F. acknowledges support from Carl Trygger's Foundation.


\begin{thebibliography}{20}
\bibitem{biasvoltage} While the general conclusion does not depend on the particular choice of biasing the DQD system, we assume here that $\mu_L=\mu+eV/2$ and $\mu_R=\mu-eV/2$ in order to simplify the discussion.
\end{thebibliography}
\end{document}